\documentclass[pdflatex]{sn-jnl}
\usepackage{preambles}
\begin{document}
\title{\Large Self-localized ultrafast pencil beam for volumetric multiphoton imaging}
\author[1,2]{\fnm{Honghao} \sur{Cao}}
\author[1,2]{\fnm{Li-Yu} \sur{Yu}}
\author[1,2]{\fnm{Kunzan} \sur{Liu}}
\author[3]{\fnm{Sarah} \sur{Spitz}}
\author[3]{\fnm{Francesca Michela} \sur{Pramotton}}
\author[3]{\fnm{Zhengyu} \sur{Zhang}}
\author[1,2]{\fnm{Federico} \sur{Presutti}}
\author[4]{\fnm{Subhash} \sur{Kulkarni}}
\author[3]{\fnm{Roger D.} \sur{Kamm}}
\author*[1,2]{\fnm{Sixian} \sur{You}}\email{\small sixian@mit.edu}
\affil[1]{\small \orgdiv{Research Laboratory of Electronics}, \orgname{Massachusetts Institute of Technology}, \orgaddress{\city{Cambridge}, \state{MA}, \country{USA}}}
\affil[2]{\small \orgdiv{Department of Electrical Engineering and Computer Science}, \orgname{Massachusetts Institute of Technology}, \orgaddress{\city{Cambridge}, \state{MA}, \country{USA}}}
\affil[3]{\small \orgdiv{Department of Biological engineering}, \orgname{Massachusetts Institute of Technology}, \orgaddress{\city{Cambridge}, \state{MA}, \country{USA}}}
\affil[4]{\small \orgdiv{Beth Israel Deaconess Medical Center}, \orgname{Harvard Medical School}, \orgaddress{\city{Boston}, \state{MA}, \country{USA}}}

\abstract{
The formation of organized optical states in multidimensional systems is crucial for understanding light-matter interaction and advancing light-shaping technologies. 
Here, we report the observation of a self-localized, ultrafast pencil beam near the critical power in a standard multimode fiber (MMF) and demonstrate its application in volumetric multiphoton imaging. 
We show that self-focusing in step-index MMFs, traditionally considered detrimental, can facilitate the formation of a nonlinear spatiotemporal localized state with a sidelobe-suppressed Bessel-like beam profile, exhibiting markedly improved stability and noise characteristics. 
By simply launching an overfilled on-axis Gaussian beam into a standard MMF, a high-quality ultrafast pencil beam can be generated through a self-localized process and readily integrated into an existing multiphoton point-scanning microscope. 
We apply this self-localized pencil beam to two-photon imaging of intact mouse enteric nervous systems, benchmarking with diffraction-limited Gaussian beams and outperforming conventional Bessel beams with reduced sidelobes and enhanced resilience to tissue-induced aberration. 
Finally, we monitor the transferrin uptake dynamics in a live human blood-brain barrier model by combining NAD(P)H/FAD‐based metabolic phenotyping with minute‐resolved 3D scans, revealing spatially and temporally resolved inter- and intra-cell heterogeneity. 
Our findings provide new insights into nonlinear dynamics of multidimensional optical systems and offer a promising approach for generating robust ultrafast pencil beams, enabling high-throughput 3D biosystem imaging to elucidate biological transport pathways and guide the design of therapeutics requiring cell-specific delivery. 
}

\maketitle
\newpage

\section{Introduction}
Wave propagation in nonlinear multimode systems enables complex wave phenomena unattainable in single-mode systems~\cite{cruz2022synthesis,wright2017spatiotemporal,wright2022physics,lu2014topological,feng2019chip,dutt2024nonlinear}.
One phenomenon that has intrigued scientists for decades is the formation of organized optical states in these multidimensional systems, including spatiotemporal solitons~\cite{malomed2005spatiotemporal,liu1999generation}, Anderson localization~\cite{schwartz2007transport,yamilov2023anderson,wiersma1997localization}, and wave thermalization~\cite{wright2016self,krupa2017spatial,zitelli2024statistics}.
These observations not only illuminate exotic physical phenomena but also hold promise for next-generation light-shaping technologies for applications such as optical manipulation~\cite{grier2003revolution}, microfabrication~\cite{faraji2021high}, and microscopy~\cite{konig2000multiphoton}. 

Among these multimode systems, high-peak-power beam shaping and pulse delivery through multimode fibers (MMFs) are inherently attractive because of the power scaling capacity and accessibility. However, realizing the full peak-power potential often proves challenging. 
Intense nonlinear pulse propagation often leads to randomization of spatial and temporal fields due to the interplay of nonlinearity and disorder.
One effective way to mitigate these randomization effects is to employ adaptive control methods by leveraging the high-dimensional degrees of freedom in multimode systems, such as input wavefront shaping through spatial light modulators (SLMs)~\cite{ploschner2015seeing,tzang2018adaptive,chen2023mitigating} or modulating pulse propagation through position-dependent perturbation~\cite{resisi2020wavefront,qiu2024spectral}. 

On the other hand, a fundamentally different paradigm from external control is offered by self-organization, wherein the nonlinear medium itself spontaneously restructures the disordered beam into an organized one under appropriate conditions, requiring minimal external shaping. 
Recent studies have highlighted self-organized beam recovery through mechanisms such as stimulated Raman scattering~\cite{baek2004single,terry2007explanation}, stimulated Brillouin scattering~\cite{lombard2006beam}, 
and Kerr beam self-cleaning~\cite{wright2016self,krupa2017spatial}. 
However, these techniques typically operate at peak powers well below the maximum power-handling potential of the media -- the critical power~\cite{marburger1975self,shen1975self}.
Approaching the critical power, self-focusing can disrupt the waveguiding condition, leading to chaotic output and potential irreversible optical damage~\cite{smith2009optical}. 
So far, self-focusing has been routinely viewed as detrimental and typically avoided in waveguide systems~\cite{seidel2018solid}. 
The possibility of achieving stable self-organization in multimode waveguides near the critical power remains largely unexplored.

In this work, we report the observation of a self-localized, ultrafast pencil beam near the critical power in a standard step-index MMF and demonstrate its utility in volumetric multiphoton imaging. 
Contrary to the assumption that self-focusing in fibers inevitably leads to catastrophic effects and optical damage, we find that the interplay between self-focusing and multimode waveguiding can yield a stable self-localized optical state (Fig.~\ref{fig-observation}a).
This localized state emerges from an incoherent superposition of symmetric fiber eigenmodes, achieving 11 dB noise suppression and enhanced robustness against external perturbations under highly nonlinear conditions. Furthermore, this ultrafast pencil beam can be simply generated by injecting an overfilled on-axis Gaussian beam into a standard MMF and readily integrated into existing multiphoton point-scanning microscopes.
Therefore, we apply this stable ultrafast pencil beam to two-photon imaging of intact mouse intestinal tissue with tdTomato-labeled enteric nervous system as a testbed.
In contrast to existing focus extension methods that rely on Bessel beams, which are constrained to low numerical aperture (NA) in practice ($\le$0.7) due to prominent sidelobes and sensitivity to aberrations, this self-localized beam achieves a pencil-like profile with negligible sidelobes in a high-NA (1.05) imaging system, exhibiting aberration resilience and efficient signal generation across a continuous 40-\textmu m axial range, all without requiring customized aperture clearance or adaptive optics. 
Finally, we monitor transferrin uptake behaviors in a live and intact human blood-brain barrier microfluidic model by combining NAD(P)H/FAD‐based metabolic phenotyping with minute‐resolved 3D scans. We find diverse and unreported uptake dynamics across endothelial cells, pericytes, and astrocytes, providing insights into biological transport pathways and the design of therapeutics requiring cell-specific delivery.



\section{Results}
\subsection{Observation of nonlinear localization in MMF}
\begin{figure}[hbtp!]
\centering
\includegraphics[width=1\textwidth]{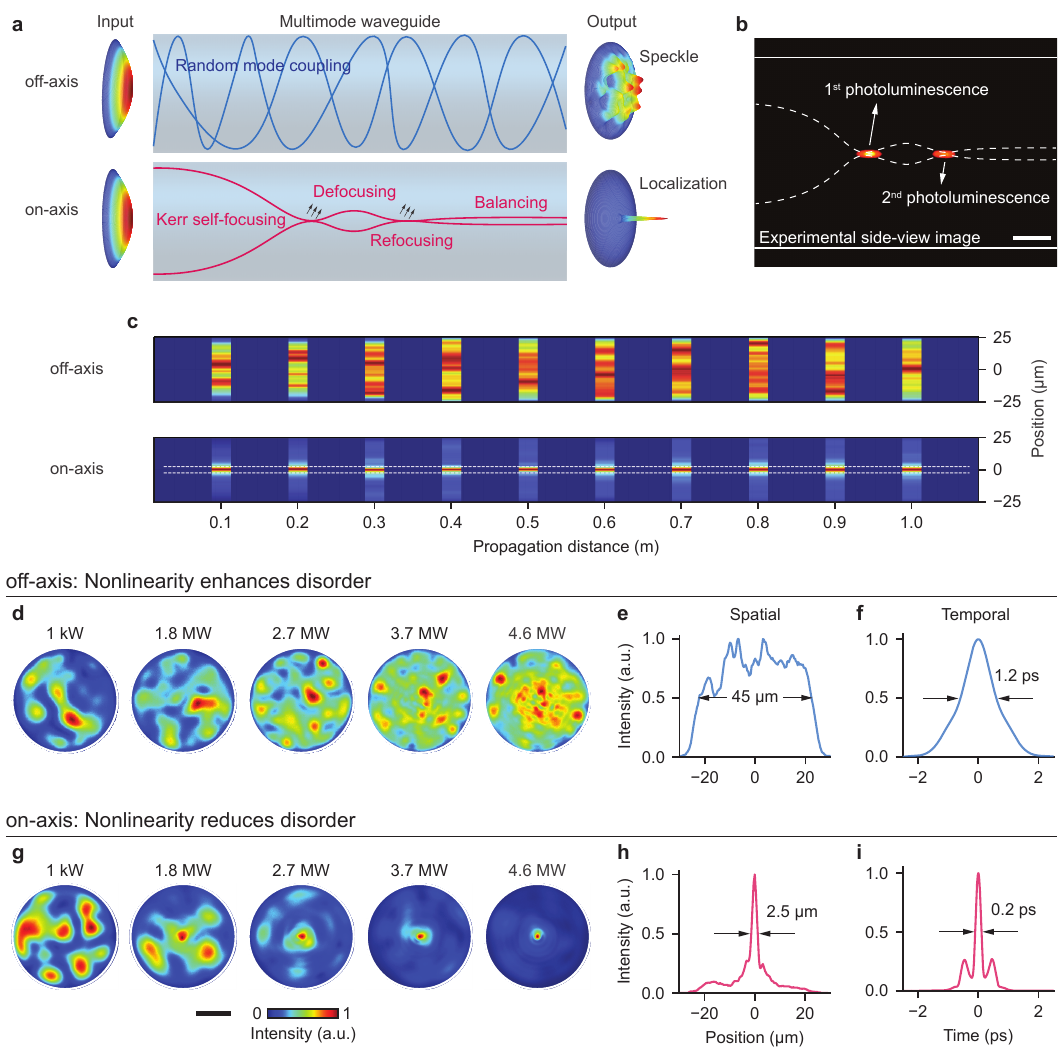}
\caption{\textbf{Experimental observation of nonlinear localization in a step-index multimode fiber.} 
\textbf{a}, Schematic of nonlinear pulse propagation in a multimode waveguide without localization (off-axis) and with localization (on-axis).
\textbf{b}, Experimental side-view image showing double photoluminescence caused by self-focusing, defocusing, and refocusing, as indicated by the dashed lines. Scale bar: 10 \textmu m.
\textbf{c}, Experimental spatial evolution of nonlinear multimodal pulse propagation. Output spatial profiles were measured at different fiber lengths, without and with localization. The side view shows the maximum intensity projection along one axis across the fiber center. The dashed lines indicate the confinement of optical power under the localization condition.
\textbf{d}--\textbf{f}, Multimodal nonlinear output without localization. Near-field images (\textbf{d}) show the output spatial distribution as input peak power increases. Horizontal-line spatial intensity profile (\textbf{e}) and temporal autocorrelation trace (\textbf{f}) at 4.6 MW show the spatial and temporal degradation.
\textbf{g}--\textbf{i}, Multimodal nonlinear output with localization. Near-field images (\textbf{g}) show the output spatial distribution as input peak power increases. Scale bar: 20 \textmu m. Horizontal-line spatial intensity profile (\textbf{h}) and temporal autocorrelation tace (\textbf{i}) at 4.6 MW show the spatial and temporal localization. A spectral bandpass filter with a central wavelength of 1025 nm and a bandwidth of 25 nm is used to select the frequency components near the pump frequency.
}
\label{fig-observation}
\end{figure}

Laser pulses with a center wavelength of 1030 nm, temporal duration of 219 fs, and a Gaussian spatial profile were launched into a standard 50-\textmu m-core silica step-index MMF (see Methods). Under the linear propagation regime with  a low input peak power (1 kW), the output spatial profile shows a speckled pattern due to random mode coupling. Under the nonlinear propagation regime with higher powers ($>$ 1 MW), the speckle patterns persist with increased granularity due to the increased nonlinearity and intermodal interactions in the MMF (Fig.~\ref{fig-observation}d--f), under non-ideal (not perfectly on-axis) input conditions, which aligns with existing observations~\cite{redding2013all,wright2015controllable,qiu2024spectral}. 

Interestingly, we observed that, when we aligned the input beam on-axis and well-centered at the fiber entrance, achieved using either a spatial light modulator or a tilt-adjusted collimation lens, the highly disordered output spatiotemporal profile under strongly nonlinear conditions (near the critical power) self-organized into a stable and centrally localized spatial profile and a shorter temporal duration (212 fs) (Fig.~\ref{fig-observation}g--i). 
As shown in Fig.~\ref{fig-observation}g, the output speckled pattern persists under the linear propagation regime (peak power 1 kW) due to random mode coupling. 
However, when we increased the peak power to megawatt levels, a distinct central peak emerged and the speckled profiles gradually transformed into a centrally localized profile near the self-focusing critical power. 
The self-organized localization and stabilization (Fig.~\ref{fig-observation}c) demonstrate that the self-focusing, when regularized by appropriate launch conditions, can overcome the intrinsic disorder in nonlinear multimodal pulse propagation. 
We note a strong contrast to the previously reported Kerr beam self-cleaning characterized by 1) the strongly nonlinear regime (MW), compared to the weakly nonlinear regime (kW) in self-cleaning~\cite{krupa2017spatial}, and 2) a centrally localized spatial distribution dominated by the highest-order mode, compared to the lowest-order (fundamental) mode dominance in self-cleaning.~\cite{pourbeyram2022direct}. 
Until now, this power regime has been primarily known for the instability caused by strong nonlinearity and complex intermodal interactions~\cite{wright2022nonlinear}.

In the experiment, the nonlinear localization was first identified by characterizing the output spatial intensity distributions with a gradual increase of the pulse peak power of an on-axis Gaussian beam input (Fig.~\ref{fig-observation}g). The launch condition for nonlinear localization is a Gaussian beam at normal or tilted incidence, precisely controlled by an SLM for the 1030-nm pump setup and by a tilt-adjusted collimation lens for the 1300-nm pump setup (see Methods). 
Different levels of self-focusing were observed across the broadened spectrum, with the strongest self-focusing effect observed near the pump center wavelength at 1030 nm, where the pulse intensity was highest. At the MMF output, the strongly localized states are experimentally separated from the rest by using a bandpass filter centered at 1025 nm.  

By launching an on-axis, overfilled Gaussian beam of 67 \textmu m in diameter, the input field predominantly excites linearly polarized (LP) rotationally symmetric modes (LP\textsubscript{0n} modes), characterized by a central intensity maximum and rotational symmetry about the fiber axis. Near the critical power, self-focusing concentrates the energy of the Gaussian input beam in a rotationally symmetric manner, ensuring a single, central intensity maximum when energy transfer dominantly occurs among these rotationally symmetric modes~\cite{fibich2000critical}. 
This is evidenced by the characterization of the spatial confinement, which shows the full width at half maximum (FWHM) of the localized state (2.5 \textmu m) is close to the MMF diffraction limit ($d = \lambda$ / 2NA = 2.3 \textmu m). The spatial profile of the nonlinear localized state is highly correlated with the highest-order rotationally symmetric mode (LP\textsubscript{0,10}) that the MMF can support (Fig.~\ref{fig-observation}h). 

In addition, the transition towards spatial localization is accompanied by temporal localization, as shown by the pulse autocorrelation measurements (Fig.~\ref{fig-observation}f,i). For highly multimode fibers ($\sim$550 modes) and a spectral bandwidth of 25 nm, pulse broadening during nonlinear multimodal propagation is mainly induced by differences in the group velocities of the propagating modes (e.g., intermodal dispersion between LP\textsubscript{01} and LP\textsubscript{04} can result in a walk-off of 1.8 ps within 30 cm of the step-index MMF). 
This temporal broadening was observed in regimes without the formation of nonlinear localization (Fig.~\ref{fig-observation}f). In contrast, with the formation of nonlinear localization near the critical power, the output pulse exhibits an FWHM of 212 fs. This measurement is close to that of a reference pulse with negligible chromatic and intermodal dispersion, obtained by exciting the LP\textsubscript{07} mode in the linear regime (see Methods). 
LP\textsubscript{07} was used as a reference here because the higher-order modes in step-index MMF can maintain single-mode propagation~\cite{ma2020propagation} and LP\textsubscript{07} has a zero-dispersion wavelength near the pump wavelength~\cite{qiu2024spectral}. 
The close agreement between the reference pulse and the localized pulse suggests that the localized state is partially non-dispersive. 
In comparison, when the beam is launched off-axis, the initial fields excite a more complex mixture of modes, including higher-order and non-symmetric eigenmodes, which have multiple intensity maxima that attract self-focusing. 
Without a dominant nonlinear attractor, the linear and nonlinear intermodal interactions remain disordered, resulting in an output with a more disordered spatial profile and longer temporal duration as well as instability (Fig.~\ref{fig-stable}). 

The symmetry-facilitated nonlinear localization is further evidenced by the observation of photoluminescence in the form of self-focusing, to defocusing, to refocusing, starting from 1.2 mm away from the input fiber facet, captured by a side-view microscope (see Methods) of the fiber with on-axis launching input (Fig.~\ref{fig-observation}b). 
The generation of the self-focused spot and enhanced photoluminescence has been reported previously~\cite{mangini2020multiphoton,ferraro2021femtosecond}, owing to self-focusing-induced MPA~\cite{polyakov2001interplay}. Note that we operated at the onset of self-focusing, with weak photoluminescence and negligible nonlinear loss. 
However, as we slightly tilted the beam from normal incidence to 0.2 and 0.4 degrees from the normal surface of the fiber input facet, the self-focusing -- defocusing -- refocusing pattern started to dissolve at 0.2 degrees, and completely disappeared at only 0.4 degrees, indicating the important role of symmetry-facilitated nonlinear localization. 

To illustrate the role of multimode waveguiding geometries in facilitating self-organized localization and propagation, we monitored the onset and stabilization of this phenomenon in fibers with different waveguiding geometries, but under similar on-axis Gaussian input. Single-mode and photonic crystal fibers constrained the beam to the fundamental mode, exhibiting instability or damage when self-focusing transferred energy into higher-order modes~\cite{seidel2018solid}. Few-mode step-index fibers enabled nonlinear localization into their highest-order mode, consistent with the trend observed in Fig.~\ref{fig-observation}. Graded-index multimode fibers exhibited early self-cleaning but quickly transitioned into unstable multiple filamentation at high powers~\cite{wright2015controllable}. Only the step-index multimode fiber supported stable, high-power localization, highlighting that a uniform refractive index profile and support for discrete, high-order symmetric modes are essential for nonlinear self-localization.
These findings demonstrate that the balance between Kerr-induced self-focusing and the multimode waveguide geometry presents a unique avenue to create spatially and temporally localized states in multimode fibers near the critical power.

\subsection{Noise-suppressed, perturbation-resistant beam delivery}
\begin{figure}[hbtp!]
\centering
\includegraphics[width=1\textwidth]{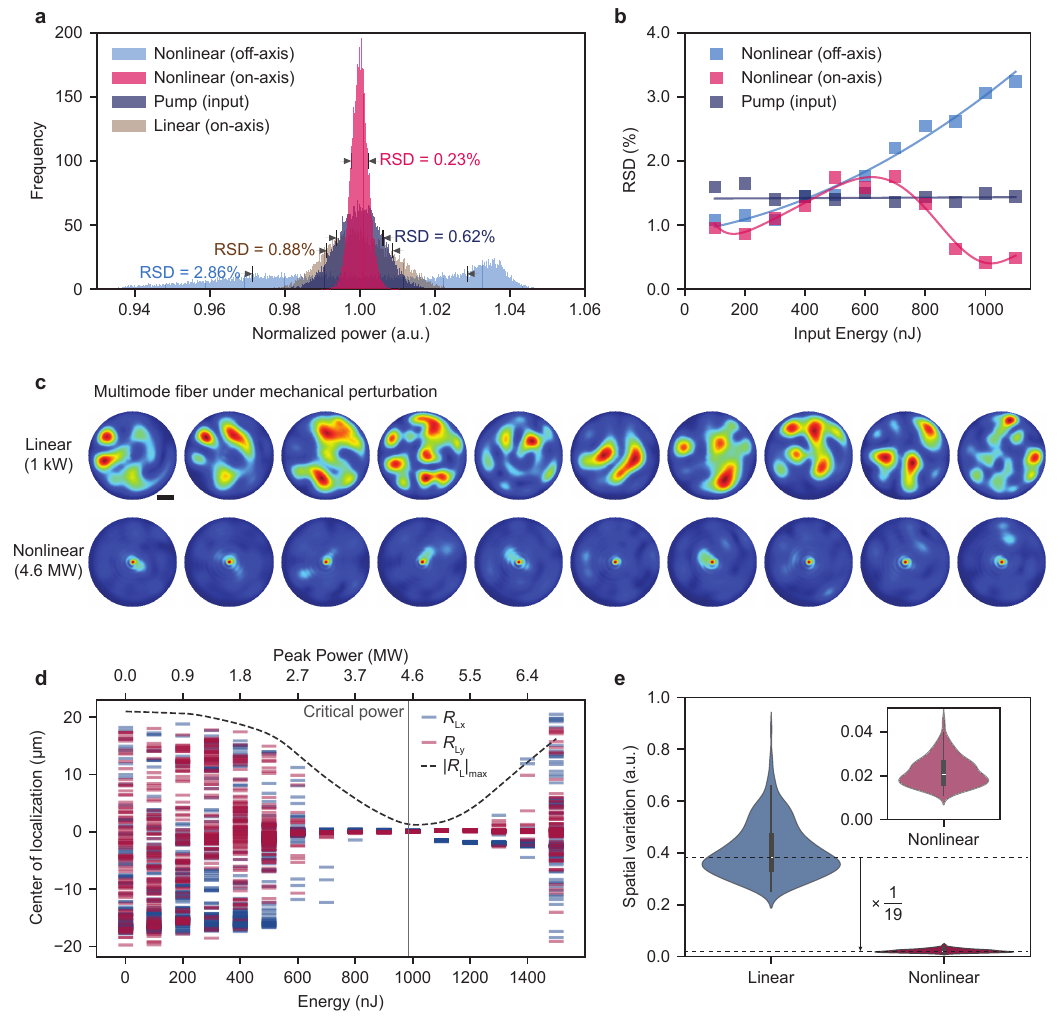}
\caption{\textbf{Stability and perturbation-resistant propagation of the nonlinear localized output.} 
\textbf{a}, Normalized power fluctuation histogram measured by a photodiode over 100 seconds, with 20,000 measurements at a time interval of 5 ms. The calculated relative standard deviation (RSD) values are shown in the plot. 
\textbf{b}, RSD values for the output without and with localization after spatial filtering as a function of increasing input energy. The RSD of the pump laser after spatial filtering is shown as a reference. The overall larger RSD values compared to \textbf{a} arise from beam position drifting.
\textbf{c}, Near-field images acquired under mechanical perturbations for both linear propagation (input peak power = 1 kW) and nonlinear localized propagation (input peak power = 4.6 MW). Identical mechanical perturbations were achieved using a motorized stage, which includes a set of 300 bending configurations. Ten representative images are displayed. Scale bar: 20 \textmu m.
\textbf{d}, Perturbation resistance of spatial localization characterized by the center of localization $\bm{R}_L = ({R_L}_x, {R_L}_y)$ under consistent bending conditions induced by a motorized stage.
\textbf{e}, Violin plots representing the distributions of spatial variation (root mean square error of the output spatial profiles to the averaged spatial profile) in \textbf{c}. Inset: zoom in on the nonlinear localized condition. 
}
\label{fig-stable}
\end{figure}
Noise is an inherent challenge in multimodal nonlinear pulse propagation, often hindering their use as reliable waveguides or light sources~\cite{rawson1983modal,kanada1983modal,rawson1980frequency}. 
Interestingly, we observed noise suppression as a result of nonlinear localization near the critical power.
We first measured the time-dependent power fluctuations under four conditions (see Methods): pump laser input, linear fiber output, nonlinear speckled fiber output (off-axis), and nonlinear localized fiber output (on-axis). The corresponding histograms are presented in Fig.~\ref{fig-stable}a. 
As expected for a linear system, the linear output exhibited a Gaussian distribution similar to that of the pump laser. The slightly larger magnitude of fluctuations in the linear output likely resulted from spatial drifting of the pump beam. 
The nonlinear speckled output displayed larger fluctuations and an irregular histogram shape. This non-Gaussian distribution with its increased noise arises from complex multimodal nonlinear effects, including four-wave mixing~\cite{bendahmane2018seeded}, transverse mode instability~\cite{wisal2024theory}, and thermal and mechanical perturbations in the environment~\cite{wada2018evaluation}. 
In contrast, the nonlinear localized output exhibited reduced power fluctuations, as quantified by the relative standard deviation (RSD) in the histograms. 
The RSD was 11 dB lower than that of the nonlinear speckled output and even 4 dB lower than that of the pump, demonstrating an unexpected noise suppression effect with the localized state in the MW-power regime in MMFs.

To understand how the noise suppression happened, we look into the possibility of intensity clamping~\cite{liu2014intensity} by measuring the output energy in the central region of the MMF core and RSD with increasing input pulse energy. 
To isolate the energy in the central region, the output beam was sent through a pinhole with an effective aperture diameter of 5 \textmu m (see Methods), allowing the measurement of the intensity clamping effect if any. 
Before the onset of localization, the output energy increases linearly with input energy. 
Near 2.7 MW, another linear increase with a two-fold steeper slope emerged due to the onset of nonlinear localization, which attracts more energy to the central region of the fiber. 
However, as the peak power approached the critical power, where the fully localized beam forms, the output energy in the central region reached a plateau, resembling the intensity clamping observed inside filaments~\cite{kandidov2011intensity}. 
Within this saturation region, the output energy remained stable despite variations in input energy. 
We observed that the fluctuations of the localized output started to decrease with increasing peak power due to intensity clamping, which is in sharp contrast with the rapidly increasing RSD of the nonlinear speckled output (Fig.~\ref{fig-stable}b). 
Near the critical power, the localized output exhibited 2–3 times lower fluctuations compared to the solid-state ytterbium laser pump, resulting from the stabilization of pump laser fluctuations via intensity clamping in the central region of the MMF.

So far, propagation of intense pulses in a MMF is fundamentally limited due to the inherent random mode coupling and sensitivity to external perturbation~\cite{horak2012multimode}. 
This causes instability and random speckle output as the pulses propagate through the MMF, which becomes more pronounced with increasing peak power due to multimodal nonlinear interactions under strongly nonlinear conditions. (Fig.~\ref{fig-observation}c,d). 
However, in our cutback experiments, when the input power approached the critical power under an on-axis Gaussian launching condition, the distance-dependent random speckle output observed in the linear regime transformed into a stable, centrally localized state despite fiber length variations, exhibiting self-trapped spatial properties within the MMF. Furthermore, we evaluated the robustness of this localized state by subjecting the fiber to controlled mechanical bending using a motorized stage~\cite{qiu2024spectral}. 
In the linear regime, the output beam profiles exhibited considerable variation with bending, whereas the nonlinear localized output remained stable (Fig.~\ref{fig-stable}c). 
To quantify how perturbation resistance evolved with increasing peak power, we computed the weighted centroid of the output profiles under consistent bending conditions across different power levels (Fig.~\ref{fig-stable}d). 
We define this spatial centroid as the center of localization (${\bm{R}_\mathrm{L}}$). 
At low peak powers, ${R_\mathrm{L}}_{x,y}$ is randomly and broadly distributed across the entire fiber core, typical for the linear regime of MMF output. 
As the input peak power increases, the centroid distribution, initially random and widespread, drastically shrinks and eventually stabilizes at the core center near the critical power. 
Beyond the critical power, this stability persists until 5.5 MW; however, at higher power (6.8 MW), instabilities arise and localization is disrupted. 
Additionally, the statistical behavior of the output beam under mechanical perturbations was analyzed for two conditions: 1) the linear regime at 1 kW, and 2) the localized regime at 4.6 MW. 
The results (Fig.~\ref{fig-stable}e) show a 19-fold reduction in spatial variation (see Methods) for the localized output, validating the improvement in perturbation resistance.

These results show that strong nonlinearity, such as self-focusing, and higher-order rotationally symmetric modes in MMFs can be leveraged together to overcome disorder and instability that are typically seen in the self-focusing regime in bulk media or in multimodal nonlinear pulse propagation in MMFs. 
In bulk media, self-focusing often leads to multiple filamentation, since an infinite continuum of wave vectors readily allows catastrophic collapse~\cite{soileau1989laser,apeksimov2016multiple}. 
By contrast, in an MMF, the beam self-focuses into a discrete mixture of LP\textsubscript{0n} modes~\cite{kibler2021discretized}, whose maximum confinement is capped by the highest-order guided mode, thus inhibiting the collapse and instability. 
Moreover, compared to linear or non-localized nonlinear propagation in MMFs -- where energy randomly couples into many transverse modes and forms unstable speckles highly sensitive to external perturbations -- the nonlinear localized state concentrates energy in the core center, reducing the influence of bending-induced index changes near the cladding~\cite{kiiveri2022refractive}. 
Simultaneously, the intense on-axis field increases the local refractive index, forming a self-induced waveguide that sustains stable, self-trapped propagation~\cite{tzortzakis2001self}. 
Therefore, the interplay between self-focusing and the multimode waveguide geometry of the MMF provides a straightforward route to exploit the high power capacity of MMFs while suppressing the speckled output, high instability, and propagation variability typically associated with the strongly nonlinear regime in MMFs.

\subsection{Volumetric multiphoton microscopy}
\begin{figure}[hbtp!]
\centering
\includegraphics[width=1\textwidth]{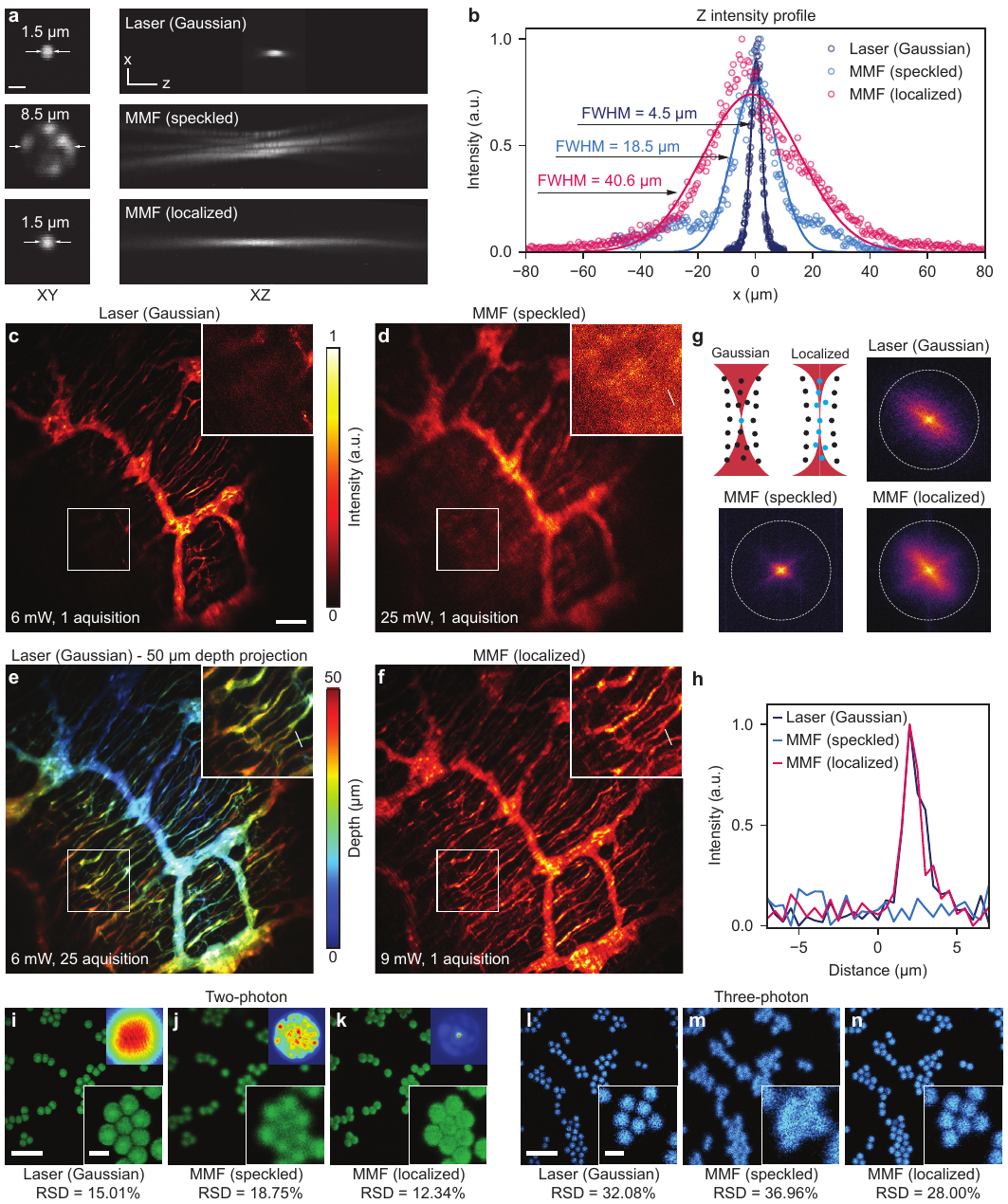}
\caption{\textbf{Volumetric multiphoton microscopy.} 
\textbf{a}, Point spread function measurement. XY scale bar: 3 \textmu m, Z scale bar: 10 \textmu m.
\textbf{c}--\textbf{f}, Two-photon fluorescence detection of the red fluorescent protein tdTomato in the small intestinal myenteric plexus of an adult Wnt1-cre:tdTomato mouse, where all the neural crest-derived cells of the enteric nervous system express tdTomato. Images were acquired with the laser Gaussian beam (\textbf{c}), laser Gaussian beam 50-\textmu m stack (\textbf{e}), and the MMF speckled (\textbf{d}) and localized (\textbf{f}) output. Scale bar: 100 \textmu m. Gamma correction = 0.8 for \textbf{c,d,f}.
\textbf{g}, Fourier spatial frequency spectra of the images in \textbf{d}--\textbf{f}, with a schematic illustrating the different focusing behaviors of the Gaussian beam and the localized beam.
\textbf{h}, Intensity profile along the line segment indicated in \textbf{d}--\textbf{f}.
\textbf{i}--\textbf{k}, Two-photon fluorescence imaging of fluorescent beads with the laser (\textbf{i}) and the MMF speckled (\textbf{j}) and localized (\textbf{k}) output. Scale bar: 40 \textmu m. Insets: (upper panel) beam profiles for each light source; (lower panel) zoomed-in views of the images. Scale bar: 10 \textmu m.
\textbf{l}--\textbf{n}, Three-photon fluorescence imaging of fluorescent beads with the laser (\textbf{l}) and the MMF speckled (\textbf{m}) and localized (\textbf{n}) output. Scale bar: 20 \textmu m. Insets: zoomed-in views of the images. Scale bar: 5 \textmu m.
}
\label{fig-volumetric}
\end{figure}

Fiber-based light delivery has gained traction in multiphoton microscopy, including \textit{in vivo} imaging in freely moving animals~\cite{zong2022large,zhao2023miniature,guan2022deep}. Compared to single-mode fibers, MMFs offer higher power threshold and potential for light shaping by supporting propagation of multiple eigenmodes including near-Bessel profiles (LP\textsubscript{0n} modes). 
Nevertheless, the use of MMFs in multiphoton microscopy has been limited due to their typically speckled and unstable output, which risks higher photodosage with lower image quality.
Here, we demonstrated that the self-localization in standard step-index MMFs can generate a pencil-like beam with minimal sidelobes, high axial extension, short pulse duration, and low intensity noise for low-photodosage and high-quality volumetric multiphoton microscopy. To investigate these advantages, we first benchmark our fiber-delivered localized output against a baseline Gaussian beam and, in the subsequent section, compare it with the LP\textsubscript{0n} near-Bessel modes to highlight its extended depth of focus and overall robustness in high-NA volumetric multiphoton imaging.

To characterize the imaging resolution, we first measured three-dimensional point spread function (PSF) profiles using 1-\textmu m fluorescent beads under three conditions (see Methods) (Fig.~\ref{fig-volumetric}a): laser input (Gaussian), non-localized MMF output (off-axis Gaussian input, speckled output), and localized MMF output (on-axis Gaussian input, self-localized output). 
In the lateral PSF, the MMF speckled output exhibited a degraded, speckled PSF, while the MMF localized output displayed a clean PSF with an FWHM close to the reference lateral PSF profile provided by the laser. 
In the axial PSF, compared to the Gaussian beam, the MMF speckled output showed a scattered random pattern due to modal-dependent foci, while the MMF localized output presented a confined, needle-like PSF. 
The comparable lateral confinement and the 10-fold axial extension  (Fig.~\ref{fig-volumetric}b) with self-organized simplicity make it a promising method for volumetric multiphoton imaging. 

Therefore, we then evaluated the spatiotemporal quality of the localized MMF output using volumetric two-photon imaging of the small intestinal myenteric plexus of an adult mouse, with the neural crest-derived cells expressing tdTomato~\cite{kulkarni2023age} (see Methods).
As expected, Gaussian beam focusing (Fig.~\ref{fig-volumetric}c) resolved the myenteric plexus with good lateral resolution but revealed only a thin section of the structure.
Without on-axis beam input to facilitate the nonlinear localization, the MMF speckled output (Fig.~\ref{fig-volumetric}d) produced blurred images with increased background, indicating degradation in both lateral and axial resolution.
In contrast, with on-axis beam input at near critical power, the MMF localized output (Fig.~\ref{fig-volumetric}f) enabled clear imaging over an extended axial range with negligible degradation of lateral resolution, thus enabling efficient single-frame acquisition of three-dimensional structural information. 
A comparison of the 50-\textmu m stack projection image with the Gaussian beam (Fig.~\ref{fig-volumetric}e) and the single-frame acquisition with the MMF localized output (Fig.~\ref{fig-volumetric}f) reveals consistent structural details.
The preservation of lateral resolution is further supported by the Fourier spatial frequency spectrum (Fig.~\ref{fig-volumetric}g) and the line intensity profile (Fig.~\ref{fig-volumetric}h).
Given the same signal yield and imaging conditions, the speckled MMF output required 25 mW while the localized MMF output required 9 mW, which was comparable to the 6 mW required by the laser but with 25 repeats of z scanning. 
The 10 times increase in the axial excitation profile provided by the localized MMF output allows 2D scanning of the Bessel-like focus to image a volume that would otherwise require 17 times higher photodosage and 25 times longer acquisition time with a conventional Gaussian focus, thereby substantially increasing imaging throughput with less photodamage~\cite{ebeling2013two}. 

Using identical imaging acquisition settings -- one pulse per pixel at a repetition rate of 1 MHz -- we compared the resolution and pulse stability of the MMF localized output and MMF speckled output against a commercial ytterbium laser as a reference (Fig.~\ref{fig-volumetric}i--n).
First, the MMF speckled output produced blurred images due to its speckled spatial profile, whereas the MMF localized output clearly resolved the microsphere edges with a resolution comparable to that of the laser.
We then extracted pixel intensity values from the central region of the beads and calculated the RSD to characterize imaging noise. 
The MMF localized output showed a lower RSD for both two-photon and three-photon imaging compared to both the MMF speckled output and the laser.
Despite the consistent trend, we note that the measured RSD in pixel values is higher than the light source power fluctuation measurements (Fig.~\ref{fig-stable}), due to additional non-laser-induced noise in our systems, such as fluorescence nonuniformity and detector electronic noise. 
Measurements that reflect more on the excitation pulse noise can be obtained by using detection methods with lower noise floor~\cite{casacio2021quantum}. 
Nevertheless, the consistently decreasing trend highlights the potential of the MMF localized output as a low-noise light source for improving the signal-to-noise ratio in situations with extremely low generation efficiency.

\subsection{Sidelobe-suppresed pencil beam}
\begin{figure}[hbtp!]
\centering
\includegraphics[width=1\textwidth]{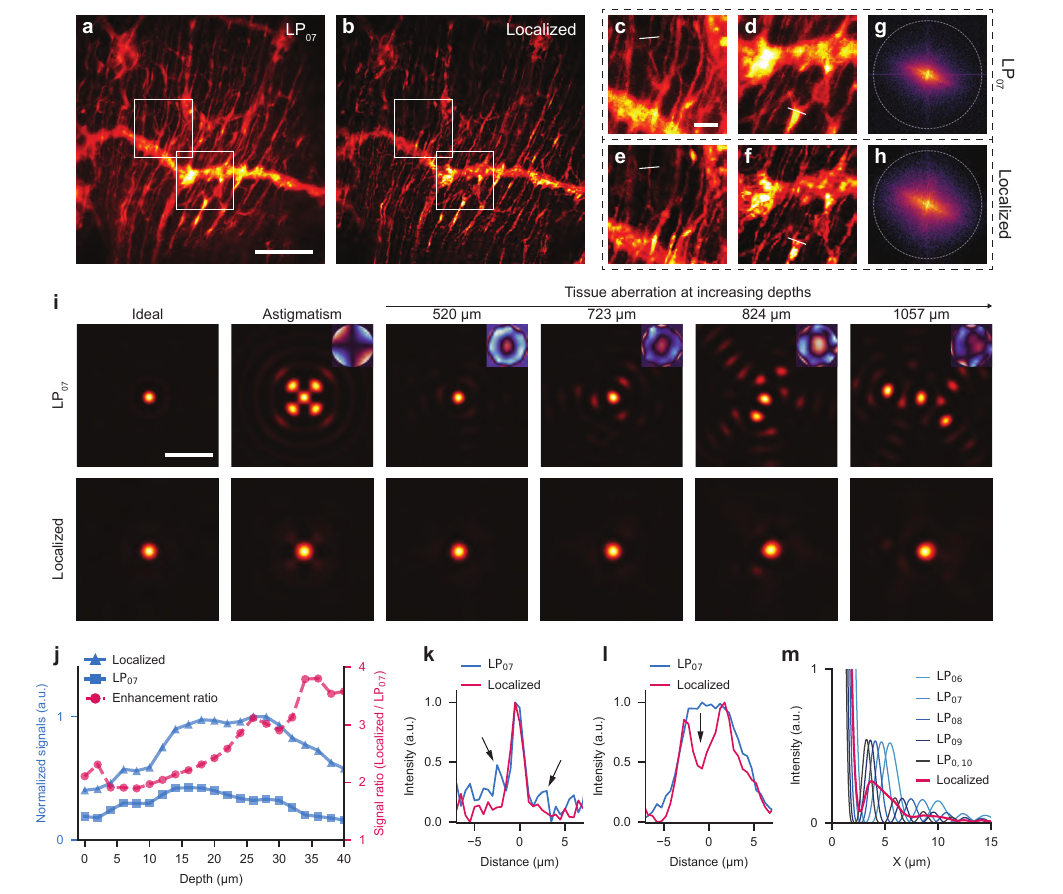}
\caption{\textbf{Volumetric imaging comparison of the single near-Bessel LP\textsubscript{0n} beam and the localized beam.} 
\textbf{a},\textbf{b}, Two-photon fluorescence imaging of the small intestinal myenteric plexus of an adult Wnt1-cre:tdTomato mouse using the near-Bessel LP\textsubscript{07} beam (\textbf{a}) and the localized beam (\textbf{b}). Scale bar: 100 \textmu m.
\textbf{c}--\textbf{f}, Zoomed-in views of the marked regions in \textbf{a},\textbf{b}. Scale bar: 20 \textmu m.
\textbf{g},\textbf{h}, Fourier spatial frequency spectra of the images in \textbf{a},\textbf{b}.
\textbf{i}, Simulated two-photon point spread functions of the near-Bessel LP\textsubscript{07} beam and the localized beam under ideal conditions, with astigmatism aberration, and with tissue aberrations. Scale bar: 3 \textmu m.
\textbf{j}, Two-photon signal levels normalized by input power for the near-Bessel LP\textsubscript{07} beam and the localized beam as a function of imaging depth. Pixels are assigned to different depths based on correlation with the Z-stack images acquired by a laser Gaussian beam.
\textbf{k},\textbf{l}, Intensity profiles along the line segments indicated in \textbf{c}--\textbf{f}.
\textbf{m}, Simulated Intensity profiles along the X-axis for various orders of the Bessel beams and the localized beam. 
}
\label{fig-aberration}
\end{figure}

Next, we investigated the imaging resolution, generation efficiency, and aberration resilience of the self-localized pencil beam in comparison to the near-Bessel LP\textsubscript{0n} beams. 
Near-Bessel beams generated by axicons, SLMs, or higher-order fiber modes are the most widely used technique for volumetric imaging to facilitate high-throughput 3D measurements in fields such as neuroscience and developmental biology~\cite{lu2017video,lu2020rapid,fan2020high}. 
While near-Bessel beams can extend the axial range beyond that of a conventional Gaussian focus, they often suffer from pronounced sidelobes and become increasingly susceptible to aberrations at higher NAs -- practical constraints that can degrade image quality. 
In free-space configurations, strategies such as Bessel droplets~\cite{george2023generation,taneja2024sidelobe,chen2024high} and incoherent pulse splitting~\cite{chen2020extended} have been explored to mitigate these limitations. Moreover, existing focus extension methods typically add system complexity, requiring dedicated optics or phase-modulating devices~\cite{chattrapiban2003generation}.
Here, we demonstrate that the ultrafast pencil beam generated by simply injecting an overfilled on-axis Gaussian input through a step-index MMF, when applied to volumetric multiphoton microscopy, exhibits sidelobe suppression, high-NA compatibility, and aberration resilience compared to the near-Bessel LP\textsubscript{0n} beams.

Experimentally, we compared two-photon imaging using the same enteric nervous system tissue site (Fig.~\ref{fig-aberration}a,b) to compare the performance of the self-localized beam with a typical near-Bessel (LP\textsubscript{07}) beam, in terms of oscillatory sidelobes (Fig.~\ref{fig-aberration}c,e,k), lateral resolution (Fig.~\ref{fig-aberration}d,f,l), and multiphoton signal generation efficiency at varying imaging depths (Fig.~\ref{fig-aberration}j).
We first show representative two-photon images acquired using the LP\textsubscript{07} mode (Fig.~\ref{fig-aberration}a) and the localized pencil beam (Fig.~\ref{fig-aberration}b). 
Fourier spatial frequency analysis (Fig.~\ref{fig-aberration}g,h) further highlights the preservation of high spatial frequencies using the localized beam. 
A quantitative comparison of two-photon signal levels as a function of imaging depth (Fig.~\ref{fig-aberration}j) demonstrates that the pencil beam maintains higher overall signal at larger depths, suggesting both improved energy confinement and more efficient excitation of deeper structures.
The lateral intensity line profiles in Fig.~\ref{fig-aberration}k,l confirm the suppressed sidelobes and improved resolving power of the localized beam compared to the LP\textsubscript{07} beam.

This enhancement in the resolution preservation, sidelobe suppression, and signal generation efficiency in thick tissues arise from the self-localized beam’s resilience to complex tissue aberration. 
To illustrate this, we model the localized beam as the incoherent superposition of a set of higher-order LP\textsubscript{0n} modes (n = 6,7,8,9,10). 
We first compare the lateral PSF under ideal conditions and in the presence of astigmatism aberration for the LP\textsubscript{07} mode and the localized beam.
To simulate tissue aberration, we applied Zernike coefficients obtained from previous work~\cite{streich2021high} in deep-tissue three-photon brain imaging to mimic realistic tissue distortions.
Under ideal conditions, both beams produce a tight central maximum, but with astigmatism and at increasing depth (520, 723, 824, 1057 \textmu m), the LP\textsubscript{07} PSF develops distorted rings, split lobes, and eventually speckled patterns. 
In contrast, the self-localized beam retains a well-confined central spot with minimal distortion (Fig.~\ref{fig-aberration}i). 
As different LP\textsubscript{0n} modes propagate along the axial direction, their sidelobes appear at varying locations and their incoherent sum can smooth out oscillations and mitigate phase errors, resulting in a robust central peak, unlike the coherent near-Bessel beam, which exhibits high sensitivity to phase aberrations (Fig.~\ref{fig-aberration}m)~\cite{mphuthi2018bessel}.
The intensity profile along the X-axis (Fig.~\ref{fig-aberration}m, consistent with the experimental result in Fig.~\ref{fig-aberration}k) and the Z-axis confirm that this incoherent combination strikes a balance of lateral confinement, sidelobe suppression, and aberration resilience.
In practice, this self-localized ultrafast pencil beam provides a simple yet robust solution for generating a high-NA, extended-depth, sidelobe-suppressed volumetric multiphoton imaging in scattering tissues, without requiring custom beam-shaping or aberration-correction optics.

\subsection{Monitoring of transferrin uptake in human blood-brain barrier model}
\begin{figure}[hbtp!]
\centering
\includegraphics[width=1\textwidth]{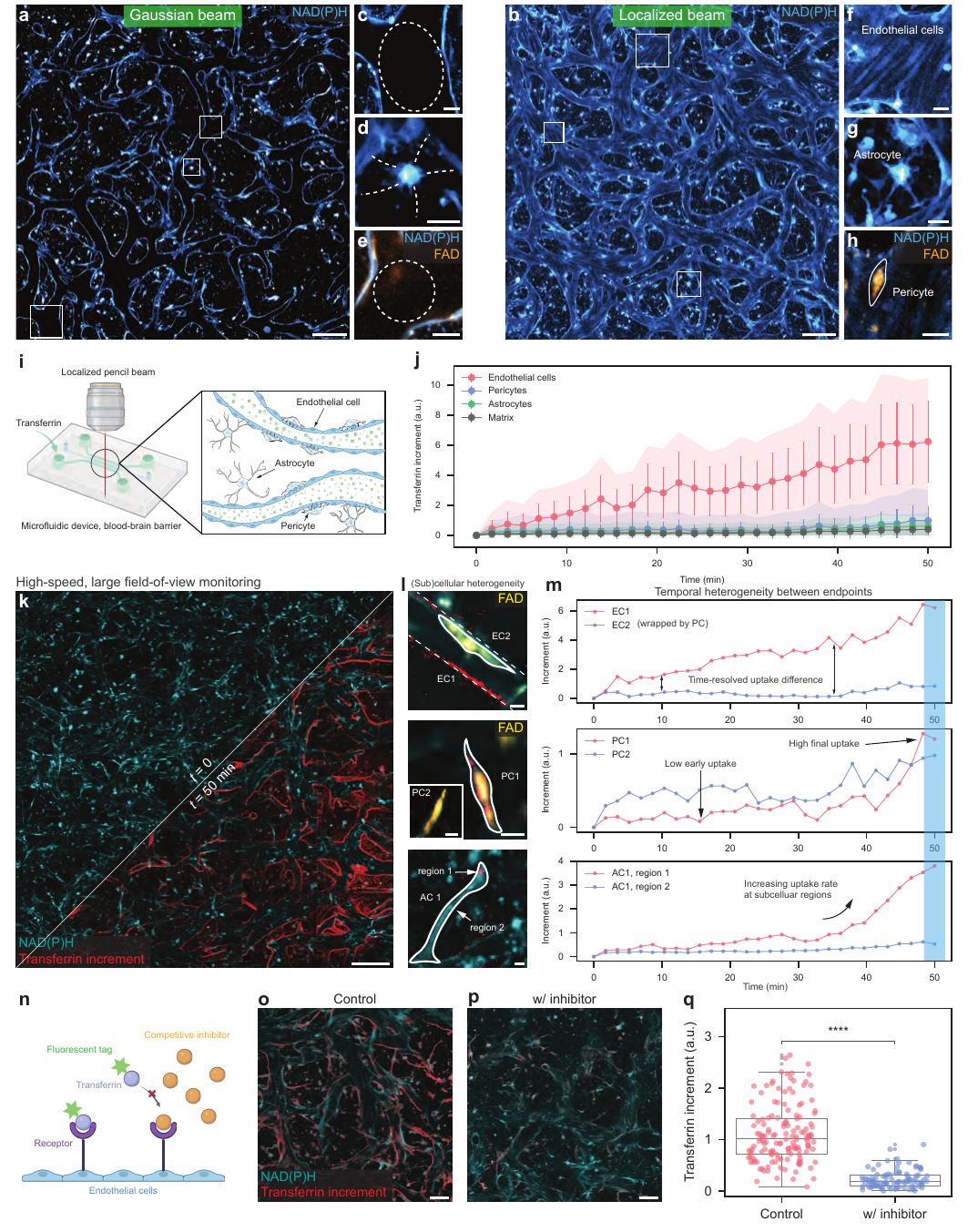}
\caption{\textbf{Monitoring transferrin uptake in a human blood–brain barrier microfluidic model.} 
\textbf{a}--\textbf{b}, Three-photon autofluorescence imaging of NAD(P)H in a fixed human blood–brain barrier model using a Gaussian beam (\textbf{a}) and a localized beam (\textbf{b}). Gamma correction = 0.6 for both images. Scale bars: 200 \textmu m.
\textbf{c}--\textbf{h}, Zoomed-in views of the marked regions. Dashed lines indicate areas missing information under Gaussian illumination due to limited axial extension. Scale bars: 30 \textmu m.
\textbf{i}, Schematic of the blood–brain barrier microfluidic device and the excitation beam configuration.
\textbf{j}, Transferrin uptake over time in endothelial cells, pericytes, astrocytes, and the gel matrix. Error bars indicate the standard deviations; shaded areas indicate the range between 5th and 95th percentiles.
\textbf{k}, Time-lapse monitoring of transferrin dynamics over 50 min. Scale bar: 200 \textmu m.
\textbf{l}, Zoom-in views of the incremental uptake of transferrin during the 50-min period. Scale bars: 10 \textmu m.
\textbf{m}, Temporal heterogeneity of transferrin uptake between endpoints. EC: endothelial cells; PC: pericytes; AC: astrocytes.
\textbf{n}, Schematic of receptor-bound transferrin on endothelial cells. Non-fluorescent transferrin was used as a competitive inhibitor to reduce binding of the labeled transferrin.
\textbf{o},\textbf{p}, Transferrin uptake after 20 min in the control group (\textbf{o}) and in the inhibitor-treated group (\textbf{p}).
\textbf{q}, Statistical comparison of transferrin uptake in endothelial cells between control and inhibitor-treated conditions.
}
\label{fig-BBB}
\end{figure}

Heterogeneous transcytosis across the blood–brain barrier (BBB) underlies both healthy neurovascular function and disease-associated barrier breakdown (e.g., in ALS or Alzheimer’s), yet direct volumetric studies of tracer kinetics in intact 3D models have been limited by the resolution–throughput trade-off of conventional microscopy. Conventional endpoint assays, such as confocal imaging of fixed, stained samples with multiple axial stacks, report only bulk tracer penetration and cannot resolve dynamic uptake patterns across the whole 3D biosystem. While three-photon NAD(P)H/FAD point-scanning can discriminate endothelial cells, pericytes, and astrocytes by their metabolic signatures in living and intact 3D BBB models~\cite{liu2024deep}, its low scan rate due to the low 3P cross section (tens of minutes per volume) precludes continuous tracer monitoring. To overcome these limitations, we applied a self-localized ultrafast pencil beam to direct, real-time monitoring of receptor-mediated protein transport at single-cell resolution across a large 3D field of view. We employed a microfluidic co-culture of human induced pluripotent stem cell (iPSC)-derived endothelial cells (ECs), pericytes (PCs), and astrocytes (ACs) embedded in a fibrin hydrogel matrix -- an in vitro BBB model that faithfully recapitulates key neurovascular morphological features and barrier functions (Methods).

First, we demonstrated that the self-localized beam maintained subcellular spatial resolution over a 2 mm × 2 mm × 50 \textmu m volume while increasing volumetric throughput by approximately 25 fold, reducing acquisition time from 25 min to 1 min (Fig.~\ref{fig-BBB}a,b). In contrast to a high-NA Gaussian focus whose limited axial extent omits critical features such as endothelial luminal membranes and astrocytic endfeet, our beam volumetrically visualized both vascular and perivascular compartments while maintaining the subcellular resolution to distinguish and track individual cells (Fig.~\ref{fig-BBB}c–h). Similar to the previously demonstrated 3P-NAD(P)H system, the simultaneous two-channel detection of NAD(P)H and FAD autofluorescence further enabled in situ calculation of optical redox ratios, revealing, for example, higher metabolic redox states in pericytes compared with endothelial cells and astrocytes~\cite{liu2024deep}.

To mimic receptor-mediated drug delivery, we perfused Alexa555-conjugated transferrin (Tf) through the microvascular networks and acquired 3D volumes every minute for 50 min until saturation in endothelial cells (Fig.~\ref{fig-BBB}i)~\cite{wiley2013transcytosis}. Initial scans (0–1 min) showed exclusive tracer confinement to the vessel lumen. Using simultaneous NAD(P)H and FAD autofluorescence to identify endothelial cells (glycolytic, NADH-rich), pericytes (oxidative, FAD-rich), and astrocytes (stellate morphology), we perfused Alexa555–conjugated transferrin and acquired volumetric datasets every minute for 50 min. Population-level curves confirmed that endothelial cells dominated tracer uptake, reaching a plateau at 45 min –- kinetics that mirror classical bulk measurements of transferrin receptor (TfR)‐mediated endocytosis (Fig.~\ref{fig-BBB}j). Uptake in pericytes remained below 30\% of endothelial cells, and astrocytes showed negligible bulk Tf accumulation, in line with static endpoint assays. 

Our high‐speed volumetric imaging with single-cell analysis then revealed that, beneath this familiar bulk behaviors, individual cells exhibit dramatically different uptake profiles (Fig.~\ref{fig-BBB}k--m), uncovering in situ heterogeneity that static or point‐scan methods cannot resolve. Some endothelial cells (e.g., EC1 in Fig.~\ref{fig-BBB}l,m) internalized transferrin within minutes, whereas adjacent endothelial cells (e.g., EC2) remained largely quiescent despite identical luminal exposure. Regions proximal to pericyte coverage correlated with suppressed endothelial uptake, consistent with pericyte-mediated suppression of vesicular transcytosis~\cite{armulik2010pericytes}. 
Although astrocytes displayed low overall transferrin accumulation, occasional subcellular “hotspots” at astrocytic endfeet produced transient fluorescence increases, indicative of localized endocytic events (Fig.~\ref{fig-BBB}l). Temporal profiling further resolved three distinct phenotypes -- “early rapid,” “delayed,” and “plateau” uptake -- for each cell class (Fig.~\ref{fig-BBB}m), a level of detail unattainable by static endpoint assays. By merging submicron, large-volume imaging with real-time kinetic readouts, our platform uncovers a previously inaccessible layer of spatial and temporal heterogeneity in BBB transport. Such spatiotemporal maps will be invaluable for dissecting barrier dysfunction in neurodegenerative diseases (e.g., ALS, Alzheimer’s), where altered permeability and uneven drug delivery remain poorly understood, and for guiding the design of brain-targeted therapeutics that must negotiate cell-specific transport barriers.

To validate the specificity of transferrin uptake in endothelial cells, we performed a competitive inhibition assay using an excess of unlabeled transferrin to block receptor-mediated binding of the fluorescent probe (Fig.~\ref{fig-BBB}n). After 20 minutes, endothelial transferrin accumulation was markedly suppressed in the inhibitor-treated group compared to controls (Fig.~\ref{fig-BBB}o,p), with quantitative analysis revealing a fivefold reduction in signal intensity (Fig.~\ref{fig-BBB}q). This workflow establishes a robust approach for validating receptor-mediated delivery mechanisms and can be readily extended to evaluate the efficiency of TfR-targeted drugs in human BBB models. This high-throughput, real-time imaging provides a scalable framework for preclinical screening of central nervous system-targeted drug delivery systems.

\section{Discussion}
In this work, we demonstrate that a simple on‐axis launch into a standard step‐index MMF, at powers near the critical self‐focusing threshold, spontaneously reorganizes into a stable, noise‐suppressed, pencil‐like ultrafast beam. This self‐localized state stands in sharp contrast to the typically observed speckled, unstable outputs at high powers, demonstrating that self‐focusing -- when operated under symmetric launch conditions -- can act as a robust mode attractor rather than a source of catastrophic collapse.

This self-localized state results from a balance between Kerr self-focusing and the MMF’s discrete modal structure, yielding a centrally concentrated, Bessel-like beam with suppressed sidelobes and sub-220~fs pulse duration. Experimentally, we show that this self‐localized state 1) delivers near-diffraction-limited lateral resolution and extends axially by 10 times even in a high-NA (1.05) imaging system and through scattering tissue; 2) exhibits reduction in power fluctuations compared to both the pump laser and a speckled MMF output ($>$ 10 dB); and 3) maintains its profile under hundreds of discrete fiber bend configurations. Together, these properties of submicron focus, extended depth of field, low noise, and mechanical robustness constitute a unique light‐delivery paradigm that requires no external adaptive optics or specialized fibers.

By integrating this self‐localized pencil beam into a standard multiphoton scanning microscope, we achieve real-time, high‐NA, large-field volumetric metabolic imaging at rates ($\sim$1~volume/min for a 1~mm~$\times$~1~mm~$\times$~50~\textmu m field) previously unattainable (vs 1~volume/25~min in previous setting~\cite{liu2024deep}), while preserving diffraction‐limited lateral resolution. When imaging tdTomato‐expressing neuronal structures in intact mouse enteric tissue, the pencil beam produced image quality and lateral resolution comparable to a diffraction‐limited Gaussian focus, yet provided an order‐of‐magnitude larger axial extension. Relative to near-Bessel beam or speckled MMF outputs, the self‐localized beam delivered higher signal ($>$ 3-fold) at larger depths with greatly reduced sidelobes and enhanced resilience to tissue‐induced aberrations. These advantages translate directly into suppressed sidelobes at high NA and in thick tissues, lower photodamage, faster 3D acquisition, and simpler optical layouts (no axicons, SLMs, or extra relay optics), collectively expanding the practical capabilities of extended-depth-of-focus multiphoton imaging.

Most importantly, we leveraged the high-resolution, high-throughput volumetric imaging to monitor transferrin uptake in a 3D human BBB microfluidic model. By combining NAD(P)H/FAD‐based metabolic phenotyping with minute‐resolved 3D scans, we mapped transferrin transport across endothelial cells, pericytes, and astrocytes in a living and intact 3D BBB model. While bulk assays confirm that endothelial cells reach tracer saturation by around one hour, our single‐cell analysis uncovered pronounced spatiotemporal heterogeneity: adjacent endothelial cells can differ by orders of magnitude in uptake kinetics, pericyte coverage locally suppresses transcytosis, and rare astrocytic ``hotspots'' exhibit subcellular endocytic events. These dynamic, spatially resolved transport signatures such as ``rapid,'' ``delayed,'' and ``nonresponsive'' phenotypes have not been accessible in prior work. Moreover, such heterogeneity is likely to be a critical factor in neurodegenerative diseases (e.g., Alzheimer’s, ALS), where subtle shifts in BBB permeability or localized barrier dysfunction may drive neuroinflammation and impair therapeutic delivery. By providing a scalable, preclinical platform for real‐time, 3D pharmacokinetic screening, this approach holds promise for accelerating the development and optimization of brain‐targeted biologics, nanoparticles, and gene‐delivery vectors.

In summary, the self‐localized ultrafast pencil beam constitutes a significant advance in multiphoton volumetric imaging and multimode nonlinear optics: it harnesses native nonlinear dynamics to generate a high‐quality, low‐noise, extended‐depth focus without external ``beam‐forming'' hardware. Its unique combination of lateral resolution, axial extension, and resilience to aberrations and perturbations opens new avenues for high-resolution, high‐throughput 3D imaging spanning neuroscience, developmental biology, and pharmacology where dynamic processes across large volumes and at submicron resolution in thick samples have previously been challenging to access. By harnessing the native nonlinear physics of multimode fibers, we believe this approach will democratize high‐NA, extended‐depth imaging, making it accessible to any laboratory with a standard multiphoton microscope, and will spur new discoveries across a wide range of biomedical fields.

\section{Methods}
\subsection{Fiber launch setups}
A mode-locked ytterbium laser (Light Conversion, Carbide) was used as the pump source, generating pulses with a central wavelength of 1030 nm, a pulse duration of 219 fs, and a repetition rate of 1 MHz. 
The pump at 1300 nm was generated using an optical parametric amplifier (OPA) (Light Conversion, Cronus-3P), delivering pulses with a duration of 46 fs at the same repetition rate of 1 MHz.
The fiber under test was a step-index MMF with a core radius of 50 \textmu m and a numerical aperture (NA) of 0.22 (Thorlabs, FG050LGA). 
The collimated laser beam (3.6 mm in diameter) was directed onto an SLM (Holoeye, Pluto) and subsequently conjugated to the fiber input facet using a 4-F relay system. 
In this system, the SLM served as the first lens (focal length = 1020 mm), while an achromatic doublet lens (focal length = 19 mm) acted as the second lens. 
This configuration resulted in a focal spot size of 67 \textmu m, overfilling the fiber core with a coupling efficiency of 55\%. 
The SLM was employed to precisely control the beam position, size, and incident angle for optimized MMF launching. 
For the LP\textsubscript{07} linear propagation reference condition, we configured the SLM as a binary phase plate to achieve LP\textsubscript{07} mode excitation~\cite{demas2015free}.
We note that the SLM is not strictly necessary for creating nonlinear localization; instead, it can be replaced by a finely tilt-adjusted collimation lens, as we demonstrated under the 1300 nm pump condition.

\subsection{Optical characterizations}
A spectral bandpass filter with a central wavelength of 1025 nm and a 25 nm bandwidth was used to select frequency components near the pump wavelength. The output spectra of the MMF were measured using a near-infrared spectrometer (Ocean Optics, NIRQuest), and the pulse duration was measured using an autocorrelator (Light Conversion, GECO) in non-collinear mode.
The near-field and far-field output spatial profiles were measured using two CMOS-based cameras (Allied Vision, Mako), positioned after a 4-F relay system and a collimation lens with an additional 4-F relay system, respectively.
Side-view images were captured by the same camera, placed after a 4-F relay system on the side of the fiber input end. A spectral bandpass filter with a central wavelength of 530 nm and a 55 nm bandwidth was added in front of the camera to select the visible photoluminescence.
The macrobending perturbation experiments were conducted using a stepper motor-based fiber shaper device~\cite{qiu2024spectral}.

\subsection{Stability assessment}
To quantify the degree of spatial localization and its stability, we introduce the concept of the center of localization (Fig.~\ref{fig-stable}d), denoted as  $\bm{R}_\mathrm{L}$. This metric represents the weighted centroid of the output spatial profile, with the weights determined by the highest-intensity pixels. It is calculated as follows:
\begin{equation} \label{center-of-localization}
    \bm{R}_\mathrm{L}(x,y) = ({R_\mathrm{L}}_x, {R_\mathrm{L}}_y) = \frac{\sum\limits_{i=1}^{N} {I_\mathrm{L}}_i (\bm{x}_i + \bm{y}_i)}{\sum\limits_{i=1}^{N} {I_\mathrm{L}}_i},
\end{equation}
where $I_\mathrm{L}$ represents the pixel intensities above the 99.9th percentile, and $x$ and $y$ are the coordinates in the transverse plane with the origin at the geometric center of the fiber cross-section. This metric allows us to assess the localization position and stability of the output spatial distributions under varying perturbations such as macrobending.

Additionally, we calculated the pixel-wise root-mean-square error (RMSE) of a set of 100 output spatial profiles under perturbations against their averaged spatial profile (Fig.~\ref{fig-stable}e):
\begin{equation} \label{spatial-variation}
    \mathrm{spatial\ variation} = \sqrt{\frac{\sum\limits_{i,j \in \mathrm{ROI}} \left(I_{ij} - \bar{I}_{ij}\right)^2}{N}}, 
\end{equation}
where $I_{ij}$ represents the intensity value at pixel $(i, j)$ in an individual spatial profile, $\bar{I}_{ij}$ represents the intensity value at pixel $(i, j)$ in the averaged spatial profile, and $N$ is the total number of pixels within the region of interest (ROI), which corresponds to the fiber core.

\subsection{Noise and intensity clamping measurement}
The pulse power noise was measured using a Germanium photodiode (Thorlabs, S132C) operating in high-bandwidth mode (100 kHz). 
A continuously variable neutral density (ND) filter (Thorlabs, NDC-50C-4M-B) was placed before the photodiode to regulate the average power across all input conditions. 
Each measurement was recorded over 100 seconds, with a time interval of 5 ms, resulting in a total of 20,000 samples.
To measure the intensity clamping in the central region of the fiber core, a 4-F relay system was used to magnify the MMF output spatial profile by 40× using a lens pair with focal lengths of 10 mm and 400 mm. A 200-\textmu m diameter pinhole was then placed at the conjugate plane of the MMF output plane to create an effective aperture of 5 \textmu m on the MMF output plane. The input laser power was measured with a Silicon photodiode (Thorlabs, S130C), and the output power after the pinhole was measured using the Germanium photodiode (Thorlabs, S132C) with a custom-written code to synchronize their readouts. The output power and the RSD values in Fig.~\ref{fig-stable}a were recorded from power meter consoles (Thorlabs, PM100D).
Spectrum measurement for intensity clamping was taken by an optical spectrum analyzer (OSA) (Yokogawa, AQ6317B) with a bandwidth from 600 nm to 1750 nm.
For imaging noise characterization, the RSD of the fluorescent bead emission signals was calculated, assuming that the fluorescent bead emission remains approximately constant. All imaging acquisition settings were kept to be the same, particularly the PMT gain, and the average signal levels were maintained approximately constant by adjusting the excitation light source power.
The coordinates of the bead centers were first identified. Then the central region of each bead was selected as the ROI, and pixel values were extracted to plot histograms and calculate the RSD as a metric for imaging noise.
The absolute RSD values include non-laser-induced noise such as fluorescence nonuniformity and detector electronic noise. However, the relative trends highlight the differences between conditions where the only variable is the light source.

\subsection{Volumetric multiphoton microscopy}
A custom-built inverted scanning multiphoton microscope was used for the imaging experiments, with a pair of galvanometer mirrors (ScannerMAX, Saturn-5 Galvo, Saturn-9 Galvo), a dichroic mirror (Thorlabs, DMLP650L), a water immersion objective (Olympus, XLPLN25XWMP2, NA = 1.05), a microscope stage (ASI, MS2000), and a photomultiplier tube (Hamamatsu, H16201) as the photodetector. For imaging samples with different emission wavelengths, corresponding bandpass filters were placed before the photomultiplier tube~\cite{qiu2024spectral,liu2024deep,han2025system}.
The laser (Gaussian) beam overfilled the objective's back aperture (diameter of 15 mm), serving as a gold standard for diffraction-limited imaging resolution. The MMF output plane was conjugated to the imaging plane with a magnification of 0.18.
We measured the near-field and far-field spatial profiles of the MMF output under different conditions and compared them with beam propagation simulations.
The PSF measurement was done with the 1-\textmu m fluorescent beads to avoid the severe photobleaching in the 0.1-\textmu m fluorescent beads and thus ensure the accurate measurement of the same bead for all light sources.
The fluorescent beads used for two-photon imaging have an excitation and emission peak at 480 nm, 520 nm. The fluorescent beads used for three-photon imaging have an excitation and emission peak at 360 nm, 450 nm. 
The tdTomato expression in the adult enteric nervous system was visualized using whole mount small intestinal tissue from a 2-month-old Wnt1-cre:tdTomato mouse, where all neural crest-derived cells of the enteric nervous system express the red fluorescent protein tdTomato.
The Zernike coefficients for tissue aberration simulation are obtained from~\cite{streich2021high}.

\subsection{Enteric nervous system sample preparation}
Adult male Wnt1-cre:tdTomato mice~\cite{kulkarni2023age} were used for this experiment. Mice were housed in a controlled environment under a 12 h light/dark cycle at 22 $\pm$ 2$^{\circ}$C, 55 $\pm$ 5\% humidity with access to food and water ad libitum. All animal procedures were carried out strictly under protocols approved by the Animal Care and Use Committee of Beth Israel Deaconess Medical Center (BIDMC) in accordance with the guidelines provided by the National Institutes of Health. Briefly, mice were euthanized, and the gastrointestinal tissues were exposed after performing a laparotomy. The small intestinal tissue was removed, and the luminal contents were flushed by using ice-cold sterile 1X phosphate-buffered saline (PBS). The tissues were then cut into 2 cm long segments and kept in ice-cold PBS in the dark till the time they were used for imaging.

\subsection{Microfluidic device fabrication}
Microfluidic devices were prepared following established soft lithography procedures.\cite{pavlou2025engineered} In summary, polydimethylsiloxane (PDMS; Sylgard\textsuperscript{\textregistered} 184 Silicone Elastomer Kit, Dow Corning) was mixed at a 10:1 base-to-curing agent ratio and poured into a multisection polycarbonate mold fabricated using computer numerical control milling. The PDMS was cured at 80\textdegree{}C for 24 hours and then sectioned into 12 distinct microfluidic units. Biopsy punches of varying diameters (Ø4 mm, Ø1.5 mm; Miltex\textsuperscript{\textregistered}) were used to create the inlets, outlets, and media reservoirs of the microfluidic devices. PDMS surfaces were cleaned with adhesive tape (Scotch\textsuperscript{\textregistered}) and sterilized by autoclaving. Devices were then bonded to glass slides (VWR, 48366-089) using air plasma treatment (Harrick Plasma; high power, 2 minutes). The glass substrates were pretreated by a 10-minute sonication in absolute ethanol followed by drying with compressed air. Bonded devices were baked at 80\textdegree{}C overnight before use.

\subsection{Microfluidic BBB culture}
Astrocytes, endothelial cells, and pericytes were derived from the human induced pluripotent stem cell line iPS11 (Alstem) based on established protocols.\cite{richards2025novel,perriot2018human,wang2020robust} To prepare cell-laden fibrin hydrogels, detached cells were resuspended in 4~U\,mL\textsuperscript{-1} thrombin (Millipore Sigma) to generate stock solutions of 22\,$\times$\,10\textsuperscript{6}\,cells\,mL\textsuperscript{-1} endothelial cells, 12.2\,$\times$\,10\textsuperscript{6}\,cells\,mL\textsuperscript{-1} astrocytes, and 1.8\,$\times$\,10\textsuperscript{6}\,cells\,mL\textsuperscript{-1} pericytes. Equal volumes of each suspension were combined to form the tri-culture mixture. For seeding, 15 \textmu L of the tri-culture was mixed with 15 \textmu L of 6\,mg\,mL\textsuperscript{-1} fibrinogen (Millipore Sigma) and immediately injected (28 \textmu L total) into each microfluidic device. Hydrogels were polymerized at 37\textdegree{}C for 15 minutes, followed by the addition of 140 \textmu L culture medium to each media channel. Interstitial flow was initiated 24 hours post-seeding by inserting syringes into one side of each device and maintained until vessels became fully perfusable. Microfluidic cultures were maintained in VascuLife\textsuperscript{\textregistered} VEGF Endothelial Medium (LifeLine Technologies) supplemented with 0.19\,U\,mL\textsuperscript{-1} heparin, 20\,ng\,mL\textsuperscript{-1} CNTF (Peprotech), and 10 \textmu M SB431542 (Selleckchem). Medium was changed every day.

\subsection{Transferrin experiments}
On day 14 of the co-culture, when the networks reached physiological cell-to-cell ratios, media was completely removed before adding 40 \textmu L of a 0.1\,mg\,mL\textsuperscript{-1} Alexa Flour 555-labeled transferrin solution (Invitrogen) to each chamber of the microfluidic device. For inhibition studies, devices were pretreated with 100 \textmu L of a 1\,mg\,mL\textsuperscript{-1} unlabeled transferrin (Jackson ImmunoResearch) solution for 30 minutes. After pretreatment, the solution was removed and replaced with 40 \textmu L of a mixture containing 1\,mg\,mL\textsuperscript{-1} unlabeled transferrin and 0.1\,mg\,mL\textsuperscript{-1} Alexa Fluor 555-labeled transferrin to initiate uptake studies.

\section{Data availability}
All data supporting the findings of this study are available in the article and its supplementary information files. The datasets acquired and analyzed in the study are available from the corresponding authors upon request. Source data are provided with this paper.

\section{Code availability}
The code used in this work is described in Methods. Example codes for data acquisition and processing are provided with this paper.

\bibliographystyle{naturemag}
\bibliography{references}

\section{Acknowledgement}
The work was supported by 
MIT startup funds,
Novo Nordisk Research Development US, Inc.,
NSF CAREER Award (2339338), 
CZI Dynamic Imaging via Chan Zuckerberg Donor Advised Fund (DAF) through the Silicon Valley Community Foundation (SVCF),
NIA R01AG66768, 
R21AG072107, 
Diacomp Foundation (Pilot award Augusta University),
and Pilot grant from the Harvard Digestive Disease Core (SK). 
H.C. and K.L. acknowledge support from the MathWorks Fellowship. 
L.Y. acknowledges support from the Claude E. Shannon Award and the MathWorks Fellowship. 
We express our sincere gratitude to Matthew Yeung and James Fujimoto for their valuable discussions and insights.
We thank Srinivas Puttapaka for helping provide the tissue samples used in our imaging experiments.

\section{Author information}
\subsection{Contributions}
H.C., L.Y., and S.Y. conceived the idea of the project. 
S.Y. supervised the research and obtained the funding. 
H.C. and L.Y. built the fiber setups and performed the fiber experiments and simulations. 
H.C. and K.L. built the imaging setups and performed the imaging experiments. 
S.S., F.M.P., Z.Z., and R.D.K. provided the human blood-brain barrier samples and insights on biological interpretation.
S.K. provided the enteric nervous system tissue samples and insights on biological interpretation.
H.C. and S.Y. wrote the manuscript with input from all authors.

\section{Ethics declarations}
\subsection{Competing interests}
The authors declare no competing interests.

\end{document}